\documentclass[aps,prl,reprint,superscriptaddress]{revtex4-1}
\usepackage{lineno}
\usepackage[dvipdfmx]{graphicx}
\usepackage{dcolumn}
\usepackage{bm}
\usepackage{siunitx} 
\usepackage{color}
\begin{document}
\preprint{}

\title{Piezomagnetic Properties in Altermagnetic MnTe}

\author{Takuya Aoyama}
 \email{aoyama@tohoku.ac.jp}
 \affiliation{%
  Department of Physics, Graduate School of Science, Tohoku University, 6-3, Aramaki Aza-Aoba, Aoba-ku, Sendai, Miyagi 980-8578, Japan.
 }%

\author{Kenya Ohgushi}
\affiliation{%
 Department of Physics, Graduate School of Science, Tohoku University, 6-3, Aramaki Aza-Aoba, Aoba-ku, Sendai, Miyagi 980-8578, Japan.
}%

\date{\today}

\begin{abstract}
  We examined the piezomagnetic effect in an antiferromagnet composed of MnTe, which is a candidate material for altermagnetism with a high critical temperature. We observed that the magnetization develops with the application of stress and revealed that the piezomagnetic coefficient $Q$ is 1.38$\times$10$^{-8}$ $\mu_{\rm B}$/MPa at 300 K. The poling-field dependence of magnetization indicates that the antiferromagnetic domain can be controlled using the piezomagnetic effect. We demonstrate that the piezomagnetic effect is suitable for detecting and controlling the broken time-reversal symmetry in altermagnets. 
\end{abstract}

\maketitle

Recently, antiferromagnets with broken time-reversal symmetry have attracted much attention~\cite{Yamaura2012, Arima2013, Nakatsuji2015, Naka2019, Hayami2019, Hayami2020, Hayami2020a, Yuan2020, Yuan2021}. 
Among them, a group without net magnetization is named altermagnets~\cite{Smejkal2022, Smejkal2022a}.
Altermagnetism is typically realized in antiferromagnets with sublattice degrees of freedom, in which the ordering of opposite spins between sublattices breaks the global time-reversal symmetry.
In altermagnets, spin current generation due to the anisotropic spin splitting in a band structure is expected, making them promising new spintronics materials~\cite{Naka2019, Hayami2019, Naka2021, Ma2021}.
To develop altermagnet-based spintronics, controlling antiferromagnetic domains is one of the key challenges because a multi-domain state generally masks the expected functionalities. 
However, unlike a ferromagnet with net magnetization, controlling antiferromagnetic domains is impossible using an external magnetic field.
An effective approach has been realized in antiferromagnets exhibiting linear magnetoelectric effects~\cite{Iyama2013}.
Nevertheless, since the control of antiferromagnetic domains through the magnetoelectric effect is limited to insulators, developing domain control methods applicable to a broader class of antiferromagnetic materials remains an important issue.

\begin{figure}[h]
    \centering
    \includegraphics[width=6cm,pagebox=cropbox,clip]{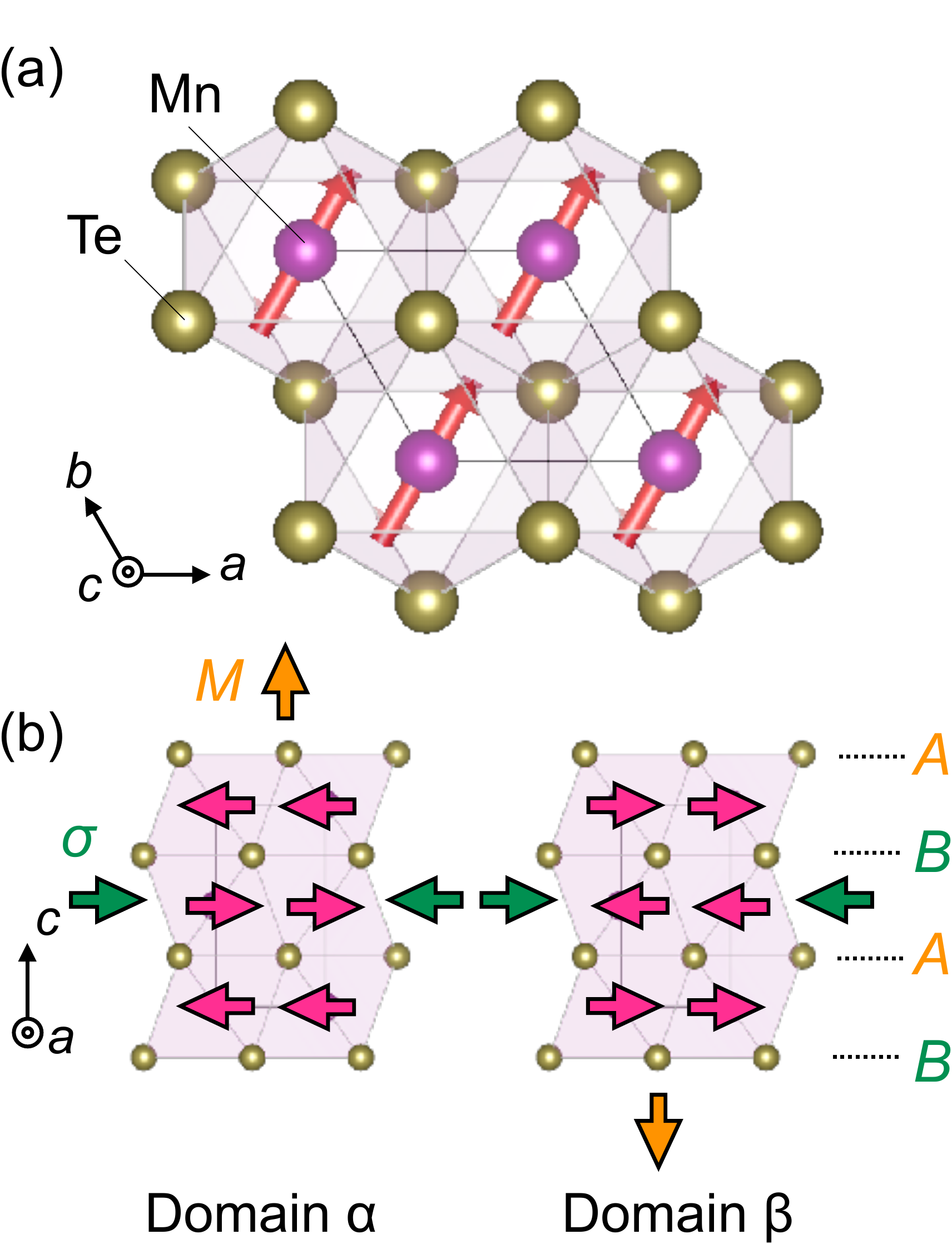}
    \caption{(a) Crystal and magnetic structure of MnTe viewed from the $c$-axis. Mn ions (purple) are in the octahedral sites between two close-packed layers of Te (gold). Red arrows represent the magnetic moments of the Mn$^{2+}$ ions. (b) Two inequivalent antiferromagnetic domains ($\alpha$ and $\beta$), which are formed below $T_{\rm N}$. Antiferromagnetic domains $\alpha$ and $\beta$ are mutually transformed by the time-reversal symmetry operation (see text for details). Therefore, the sign of the piezomagnetic tensor ($Q_{ijk}$) is opposite in between, so the application of stress ($\sigma$) produces a net magnetization ($M$) in the opposite direction. The $A$ and $B$ represent the closed-packed layers of Te. }
    \label{structure}
\end{figure}

The target material of this study is MnTe with the NiAs structure (space group of $P6_3/mmc$), which has been proposed as a typical example of an altermagnet~\cite{Smejkal2022, Smejkal2022a}. 
MnTe exhibits an antiferromagnetic transition at the N\'{e}el temperature ($T_{\rm N}$) of 307 K, below which Mn$^{2+}$ spins pointing in the $\langle$2$\bar{1}\bar{1}$0$\rangle$ direction order ferromagnetically in the $c$-plane and antiferromagnetically along the $c$-axis (Fig. \ref{structure}(a))~\cite{Komatsubara1963, Szuszkiewicz2005, GonzalezBetancourt2023, Mazin2023}. 
The altermagnetism in MnTe is well understood by considering the structural features.
The NiAs structure consists of a hexagonal close-packed lattice of anions, which involves the $ABAB$ stacking of triangular lattices.
The cations occupy the octahedral sites between the anion layers, forming a simple hexagonal lattice. 
Since the octahedral sites between the $AB$ and $BA$ layers are inequivalent, sublattice degrees of freedom appear in the cation sites.
If magnetic moments with opposite directions are aligned on these two sites, the global time-reversal symmetry is broken; this is because the spin reversal cannot be achieved by a single symmetry operation on the crystal lattice, such as a rotation and translation.
Similar considerations apply to other altermagnets, such as RuO$_2$, MnF$_2$, and CrSb~\cite{Smejkal2020, Yuan2020, Smejkal2022a}, which possess the NiAs derivative structure.

\begin{figure}[h]
    \includegraphics[width=7cm,pagebox=cropbox,clip]{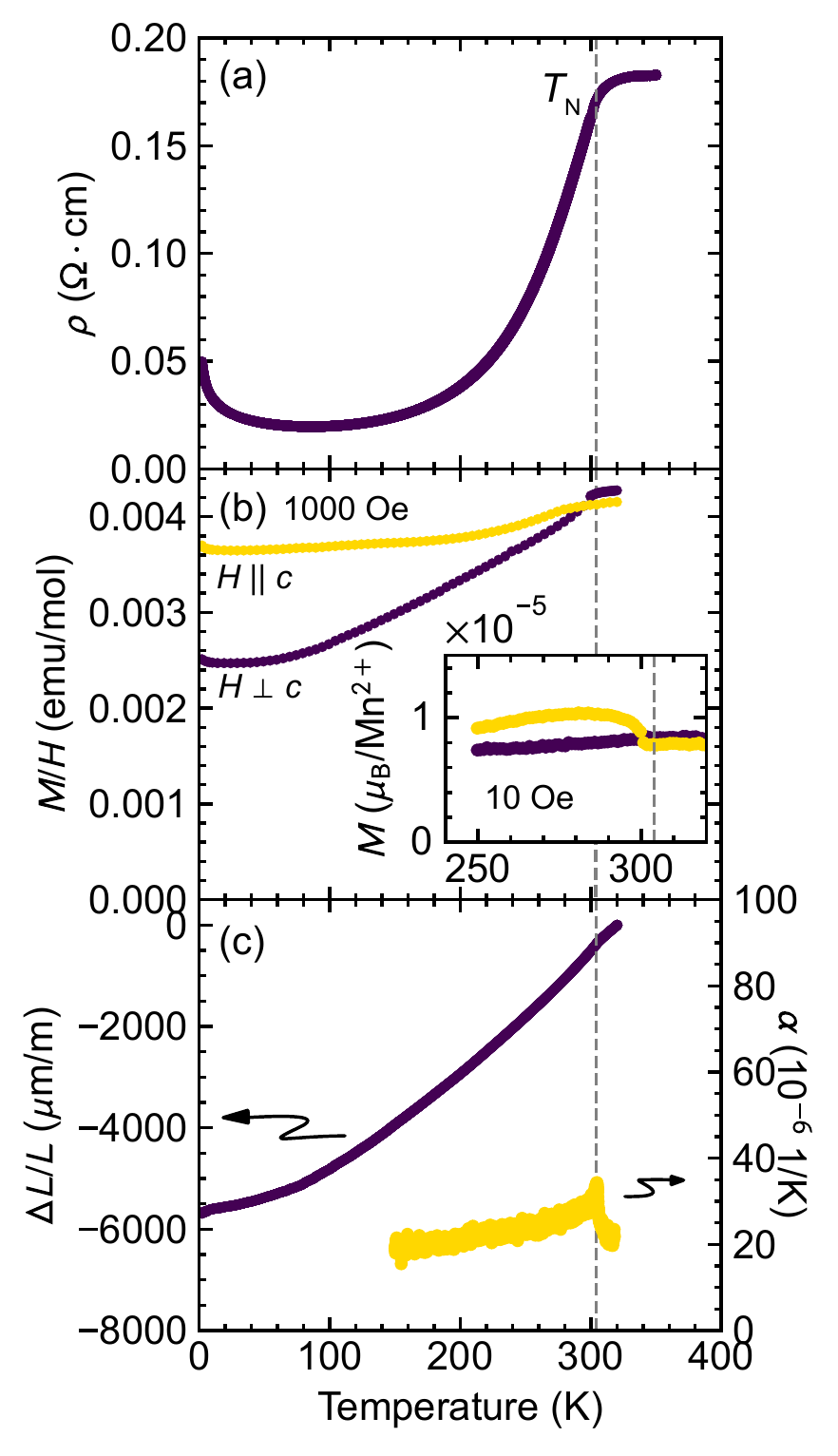}
    \caption{Temperature dependence of (a) the $c$-plane electrical resistivity ($\rho$), (b) the magnetization divided by the magnetic field $M/H$, and (c) the strain ($\frac{\Delta L}{L}$) and thermal expansion coefficient ($\alpha$ = $\partial (\frac{\Delta L}{L}) / \partial T$) in the $c$-plane. Inset of (b) represents the magnetization taken at $H$ = 10 Oe in a heating process, after cooling from the paramagnetic phase at $H$ = 10000 Oe.}
    \label{Tdep}
\end{figure}

To evaluate altermagnetism in MnTe, we focused on the piezomagnetic effect.
The piezomagnetic effect, a linear coupling between magnetization and stress, is allowed when the free energy includes a term of $F$ = $Q_{ijk}H_i\sigma_{jk}$~\cite{Moriya1959, Borovik-Romanov1960, Dzialoshinskii1958, Hayami2018, Yatsushiro2021} ($H$ being the magnetic field and $\sigma$ being the stress tensor). 
Here, the piezomagnetic tensor $Q_{ijk}$ is a third-rank axial-$c$ tensor and becomes finite when the time-reversal symmetry is globally broken, which is a useful probe for detecting the altermagnetism~\cite{Arima2013, Nakajima2015, Ikhlas2022}.
The magnetic point group in the antiferromagnetic phase of MnAs is $mmm$, meaning that $Q_{xyz}$, $Q_{yzx}$, and $Q_{zxy}$ are finite ($x$, $y$, and $z$ being [2$\bar{1}\bar{1}$0], [01$\bar{1}$0], and [0001], respectively). 
Another aspect of the piezomagnetic effect is that it may enable the control of the antiferromagnetic domains.
The antiferromagnetic domains discussed here are not 120$^{\circ}$ domains associated with three-fold rotational symmetry breaking, but two domains associated with broken time-reversal symmetry that are unique to altermagnets, as shown in Fig. \ref{structure}(b).
These two domains related to breaking the time-reversal symmetry have different signs of $Q_{ijk}$ in the free energy $F=$ $Q_{ijk}\sigma_{ij}H_k$.
Hence, an energy difference between the two domains can be induced by applying $\sigma$ and $H$ in the appropriate directions. 
Particularly, applying $\sigma$ and $H$ during cooling across the $T_{\rm N}$ is an efficient way to achieve a single-domain state.

In this study, we examined the piezomagnetic effect in MnTe by measuring the stress dependence of magnetization.
We revealed that the piezomagnetic tensor is as large as 2.5$\times$10$^{-8}$ emu/MPa at 300 K.
Controlling the antiferromagnetic domains using the piezomagnetic effect was demonstrated by changing the poling field in a cooling process across $T_{\rm N}$.

The samples used in this study were polycrystals purchased from Kojundo Chemical Lab. Co., Ltd. (lump, 99.9\% purity) and single crystals grown by chemical vapor transport using I$_2$ as an agent gas~\cite{Wasscher1969, DeMelo1990, Reig2001}.
The temperature ($T$) dependence of electrical resistivity ($\rho$) was measured in a $^4$He cryostat using the standard four-terminal method.
A commercially available SQUID magnetometer was used for magnetization ($M$) measurements under selected $H$.
Strain ($\frac{\Delta L}{L}$) was measured using the active-dummy method with a commercial strain gauge (Kyowa KFL-02-120-C1-11), and the thermal expansion coefficient ($\alpha$) was derived by differentiating the obtained $\Delta L/L$ by $T$.
The piezomagnetic effect was evaluated by measuring the magnetization under constant $\sigma$.
The sample was inserted into a piston cylinder made of an engineering plastic and pressurized using a hydraulic press for the stress application, which was applied in a parallel direction to the magnetic field. 
The stress was maintained using a clamp cell made of CuBe, and the generated stress was calculated from pressure gauge measurements. 
The magnetization of the cell was estimated to be approximately 1$\times$10$^{-6}$ emu, which is much smaller than the sample signal.

Figure \ref{Tdep}(a) shows the $T$ dependence of $\rho$ in the $c$-plane of single crystalline MnTe. 
The metallic behavior of $\rho$ over a wide $T$ range is consistent with the previous results for the bulk and thin film MnTe~\cite{Wasscher1969, Kriegner2016, Yang2016}.
However, several reports indicate an insulating behavior~\cite{Yang2016, Mori2018}. 
These discrepancies suggest that MnTe significantly changes its electronic state depending on the chemical composition and epitaxial strain.
The sharp decrease in $\rho$ at 307 K is attributed to the decrease in magnetic scattering due to the antiferromagnetic transition.
The small upturns in $\rho$ below 50 K are likely a result of localization phenomena.

Figure \ref{Tdep}(b) shows the $T$ dependence of $M$/$H$ observed under $H$ = 1000 Oe (Fig. \ref{Tdep}(b)).
Upon cooling below $T_{\rm N}$, $M/H$ with $H$ $\perp$ $c$ decreases, and $M/H$ with $H$ $||$ $c$ remains approximately constant, indicating that the magnetic moment is lying in the $c$-plane.
This agrees with previous neutron scattering and magnetic torque studies~\cite{Kunitomi1964, Komatsubara1963}.
Moreover, the magnetization measurement at a weak magnetic field of $H$ = 10 Oe reveals a parasitic magnetization on the order of 1$\times$ 10$^{-6}$ $\mu_{\rm B}$/Mn along the $c$-axis (inset of Fig. \ref{Tdep}(b)).
Since the magnetic point group $mmm$ strictly forbids the appearance of ferromagnetic components~\cite{Yatsushiro2021}, this is likely observed because of a tiny excess or deficiency of Mn$^{2+}$ ions. 
Another scenario is that the actual magnetic point group becomes much lower than $mmm$ owing to the lattice deformations.

Figure \ref{Tdep}(c) shows the strain and thermal expansion coefficient in the $c$-plane of single crystalline MnTe.
The lattice constant in the $c$-plane decreases upon cooling; the decreasing tendency enhances below $T_{\rm N}$, as observed in the $T$ dependence of $\alpha$.
This implies that the ferromagnetic interactions are favored in the shortened bonds.
The presence of strong spin-lattice coupling should produce a significant piezomagnetic effect.

\begin{figure}
\includegraphics[width=8cm,pagebox=cropbox,clip]{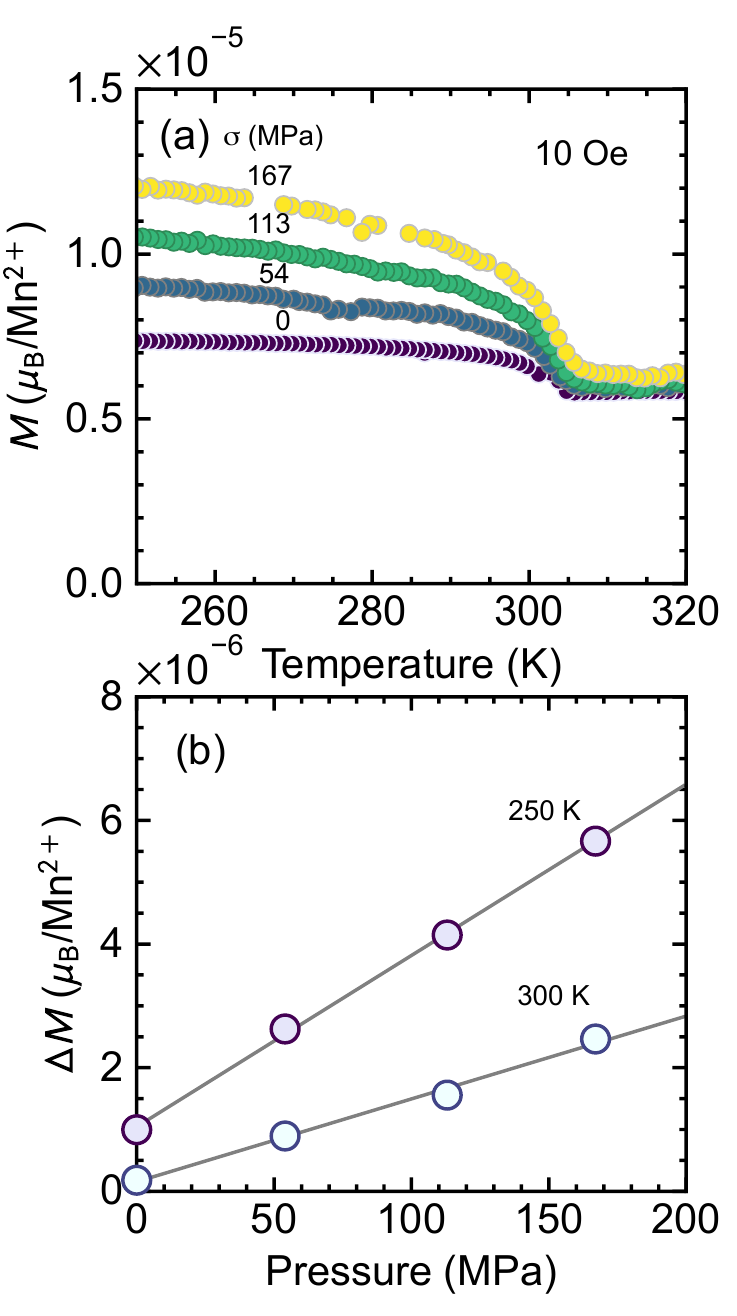}
\caption{(a) Temperature dependence of magnetization (M) for polycrystalline MnTe measured under various stresses. Measurements were performed at $H$ = 10 Oe during a heating process, after cooling from 320 K to 250 K at $H$ = 10 Oe. (b) Stress dependence of $\Delta M$ at 250 and 300 K, where $\Delta M$ represents the relative change in magnetization from 320 K. The $\Delta M$ value includes no contributions from the paramagnetic susceptibility.}
\label{PMeffect}
\end{figure}

Figure \ref{PMeffect}(a) shows the $T$ dependence of the magnetization measured under various stresses. 
These data were collected during a heating process at $H$ = 10 Oe after cooling at 10 Oe from a paramagnetic phase.
The spontaneous magnetization develops below $T_{\rm N}$ at all stresses measured and increases monotonically with increasing stress.
To evaluate the systematic change in magnetization, we plot $\Delta M$ at the fixed temperature against the stress in Fig. 2(b), where $\Delta M$ is the relative change in magnetization from 320 K.
The linear development of magnetization with respect to stress is direct evidence of the piezomagnetic effect.
Fitting with the linear function $M$ = $Q_{\rm ave} \sigma$ shown by the solid gray line yields $Q_{\rm ave}$ = 2.68$\times$10$^{-8}$ $\mu_{\rm B}$/Mn/MPa at 250 K and 1.38$\times$10$^{-8}$ $\mu_{\rm B}$/Mn/MPa at 300 K.
Here, $Q_{\rm ave}$ is the powder-averaged piezomagnetic tensor and has a relationship of $Q_{\rm ave}$ = $\frac{2}{3\pi^2}$($Q_{xyz}$ + $Q_{yzx}$ + $Q_{zxy}$) in MnTe.
This value is comparable to the piezomagnetic effect tensor $Q_{\rm ave}$ = 6.1$\times$10$^{-8}$ $\mu_{\rm B}$/Mn/MPa for MnF$_2$, which also contains Mn$^{2+}$ ions~\cite{Borovik-Romanov1960, Komuro2023}.

In altermagnets, two types of domains associated with breaking the time-reversal symmetry are formed by magnetic ordering, which are shown in Fig. \ref{structure}(b).
Herein, we demonstrated domain control using the piezomagnetic effect.
We first cooled the samples from the paramagnetic state at 330 K to the antiferromagnetic state at 250 K with a cooling rate of 10 K/$min$ under various poling fields ($H_{\rm pol}$), where we fixed the stress condition of $\sigma$ = 167 MPa.
We then measured the magnetization in the heating process under $H_{\rm meas}$ = 10 Oe.

As a result, a significant $H_{\rm pol}$ dependence of the temperature evolution of magnetization was observed, as shown in Fig. \ref{PMeffect} (a).
The spontaneous magnetization below $T_{\rm N}$ decreases as $H_{\rm pol}$ decreases from 10000 Oe to -10000 Oe.
The magnetization becomes negative when a negative $H_{\rm pol}$ is applied, even though $H_{\rm meas}$ is positive. 
The observed $H_{\rm pol}$ dependence is understood as the population of antiferromagnetic domain changes as a function of $H_{\rm pol}$ according to the mechanism described above.

\begin{figure}
    \includegraphics[width=8cm,pagebox=cropbox,clip]{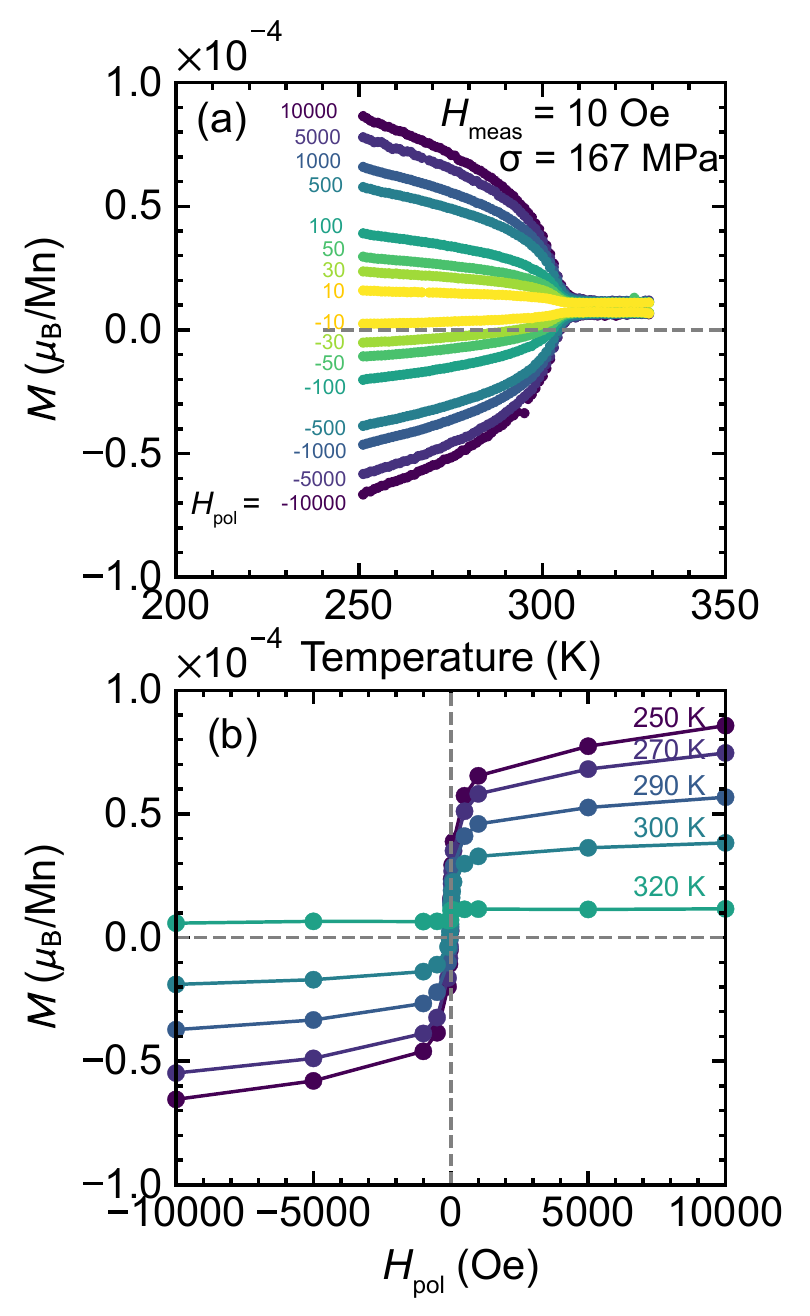}
    \caption{(a) Poling field ($H_{\rm pol}$) dependence of magnetization at $\sigma$ = 167 MPa and $H_{\rm meas}$ = 10 Oe. After cooling from 330 to 250 K at $\sigma$ = 167 MPa and $H_{\rm meas}$ = 10 Oe, measurements were performed during a heating process at $H_{\rm meas}$ = 10 Oe. (b) Poling-field dependence of magnetization at various temperatures.}
    \label{Domain}
\end{figure}

To evaluate the poling field required to achieve the single-domain state, the data sets of magnetization at several temperatures were plotted against the poling field, as shown in Fig. \ref{PMeffect}(b).
The $M$-$H_{\rm pol}$ curve saturating at as small as $H_{\rm pol}$ = 1000 Oe indicates that the antiferromagnetic domain of MnTe is highly controllable.
This is likely related to the isotropic nature of the Mn$^{2+}$ spins with quenched orbital angular momentum; the energy barrier between the two domains is relatively small. 
Another important aspect is that the $H_{\rm pol}$ required for magnetization saturation is approximately independent of temperature.
This means that the antiferromagnetic domain population depends only on $H_{\rm pol}$ across $T_{\rm N}$, and the single-domain state is robust against temperature variation after it is formed.

Notably, our observations of the piezomagnetic effect substantiate that MnTe is an altermagnet. 
Next, devices must be fabricated that demonstrate novel phenomena expected in altermagnetism, such as spin current generation.
This may be possible without great difficulties since thin films can be easily fabricated by chemical vapor transport and because $T_{\rm N}$ is above room temperature. 
However, it should be noted that net magnetization can be induced by external stress or strain from a substrate, as presented in this study. 
Therefore, when measuring novel physical properties caused by altermagnetism, it is necessary to avoid the contributions from the net magnetization induced by the piezomagnetic effect.

In conclusion, we investigated the piezomagnetic effect in the MnTe altermagnet and observed a stress-induced magnetization below the antiferromagnetic transition temperature of 307 K.
The piezomagnetic coefficient was estimated to be $Q_{\rm ave}$ = 1.38$\times$10$^{-8}$ $\mu_{\rm B}$/Mn at 300 K.
We also demonstrated antiferromagnetic domain control using the piezomagnetic effect by evaluating the poling-field dependence of the magnetization.
The poling field of $H_{\rm pol}$ = 1000 Oe is sufficient to obtain the single-domain state at $\sigma$ = 167 MPa. 

\begin{acknowledgments}
        We would like to thank T. Sato, S. Soma, T. Osumi, and K. Yamauchi for the fruitful discussions. 
        We also thank Edanz (https://jp.edanz.com/ac) for editing a draft of this manuscript. 
        This work was financially supported by JSPS KAKENHI Nos.
  JP22H00102, 
  JP19H05823, 
  and JP19H05822, 
\end{acknowledgments}

\bibliography{MnTe}

\end{document}